# Direct growth of single- and few-layer MoS$_2$ on h-BN with preferred relative rotation angles


*Aiming Yan[†, §, #], Jairo Velasco, Jr. [†, #], Salman Kahn[†], Kenji Watanabe[⊥], Takashi Taniguchi[⊥], Feng Wang[†, §, #], Michael F. Crommie[†, §, #], Alex Zettl[†, §, #,*]*

[†]Department of Physics, University of California, Berkeley, CA 94720, USA

[§]Materials Sciences Division, Lawrence Berkeley National Laboratory, Berkeley, CA 94720, USA

[⊥]National Institute for Materials Science, 1-1 Namiki, Tsukuba, 305-0044, Japan

[#]Kavli Energy NanoSciences Institute at the University of California, Berkeley and the Lawrence Berkeley National Laboratory, Berkeley, CA 94720, USA



**ABSTRACT:**

**Monolayer molybdenum disulphide (MoS$_2$) is a promising two-dimensional direct-bandgap semiconductor with potential applications in atomically thin and flexible electronics. An attractive insulating substrate or mate for MoS$_2$ (and related materials such as graphene) is hexagonal boron nitride (h-BN). Stacked heterostructures of MoS$_2$ and h-BN have been produced by manual transfer methods, but a more efficient and scalable assembly method is needed. Here we demonstrate the direct growth of single- and few-layer MoS$_2$ on h-BN by chemical vapor deposition (CVD) method, which is scalable with suitably structured substrates. The growth mechanisms for single-layer and few-layer samples are found to be**




distinct, and for single-layer samples low relative rotation angles (<5º) between the $MoS_2$ and h-BN lattices prevail. Moreover, $MoS_2$ directly grown on h-BN maintains its intrinsic 1.89 eV bandgap. Our CVD synthesis method presents an important advancement towards controllable and scalable $MoS_2$ based electronic devices.

**KEY WORDS:** molybdenum disulfide, chemical vapor deposition, heterostructure, hexagonal boron nitride, screw-dislocation driven growth, transition metal dichalcogenides

The realization of single-layer graphene on an insulating substrate[1] sparked renewed interest in van der Waals (vdW) bonded two-dimensional (2-D) materials including the exploration of new phenomena and potential applications. Transition metal dichalcogenides (TMDs) are well-known vdW 2D structures that can also be exfoliated in single atomic layer form onto insulating substrates. Notably, TMDs display many physical properties distinct from those of graphene. $MoS_2$ is a particularly noteworthy TMD in that it displays a direct electronic bandgap of 1.89 eV in single layer form and a smaller indirect gap for multi-layers. This transition allows much enhanced quantum yield of photoluminescence. Single-layer $MoS_2$ – based field effect transistors (FETs) exhibit high on/off ratio[2], and control of valley polarization and coherence[3]. These properties establish $MoS_2$ as a promising candidate for flexible electronic, optoelectronic, and photonic applications.

Although for some applications suspended bare sheets of $MoS_2$ or other 2D materials is useful, in general the monolayers (or few layers) are mated to a substrate, either for mechanical stability or enhanced processibility, or to create a desirable electronic/optical heterostructure. The mate is often a 2D vdW material itself, and fabricating heterostructures comprised of



different 2-D layered materials is a versatile approach that can integrate materials with different properties and realize new device functionalities[4–6]. Mating can be achieved through the manual transfer of individual 2-D layered materials or the direct growth of one type of 2-D material on top of or adjacent to another[7,8]. Although the transfer method enables virtually any combination of layered materials in a heterostructure, it is tedious and effectively non-scalable. Direct transfer can also trap impurities or residues at the interface between individual layers during the transfer [6,9]. In contrast, direct growth of a 2-D layered material on top of or next to another is a more scalable and controllable method and yields clean interfaces[7,8]. Electrically insulating 2-D h-BN has been shown to be a superior substrate to $SiO_2/Si$ for graphene electronic devices due to the flat surface of h-BN and less charge inhomogeneity[10–12]. Similarly, it has also been shown that $MoS_2$ transferred onto h-BN exhibits excellent device quality[13,14]. The direct growth of $MoS_2$ on h-BN by CVD methods would be an important advance in fabricating high-quality $MoS_2$ electronic devices in a scalable and controllable way.

CVD growth of $MoS_2$ on different substrates has been investigated extensively in the past two years[15,16]. With the seeding method, monolayer $MoS_2$ can be grown on various substrates[17,18] including h-BN.[18] However, growing $MoS_2$ directly on h-BN without any seeding method yields a clean interface between as-grown $MoS_2$ and h-BN substrate, which realizes the direct mating of these two layered materials. This not only allows the fabrication of higher-quality devices without extensive annealing processes, but also promotes interesting physics due to the direct coupling of $MoS_2$ and h-BN lattices. Here we demonstrate that single- and few-layer $MoS_2$ can be grown directly on exfoliated high-quality h-BN flakes using CVD. We find that the nominal growth mechanisms are different for single-layer and few-layer $MoS_2$. Single-layer samples



display low relative rotation angles (<5°, with the specific definition for relative rotation angle discussed below) between the MoS$_2$ and h-BN hexagonal lattices.

Fig. 1(a) shows the two-zone furnace setup for CVD growth of single- and few-layer MoS$_2$ on exfoliated h-BN on SiO$_2$/Si substrates. Unlike the reported one-zone furnace setup for the CVD growth of single-layer MoS$_2$ on bare SiO$_2$/Si substrates[15,16], a two-zone furnace allows the separate control of S and MoO$_3$ sources and enables greater tunablility of the reaction process. The growth of single-layer MoS$_2$ on exfoliated h-BN is shown in the schematic in Fig. 1(b), where the green colored flake represents h-BN exfoliated on a SiO$_2$/Si substrate and the isolated blue polygons represents single-layer MoS$_2$ flakes grown on the h-BN. Fig. 1(c) is an optical image that depicts the typical end result of such a growth. The light green h-BN flakes are typically 20×10 μm$^2$ and the dark green MoS$_2$ islands have typical size 2-3 μm and are scattered randomly on the h-BN flake. On occasion, we also observe MoS$_2$ islands with higher optical contrast, suggesting growth of MoS$_2$ with layer number >1 is also possible (see below).

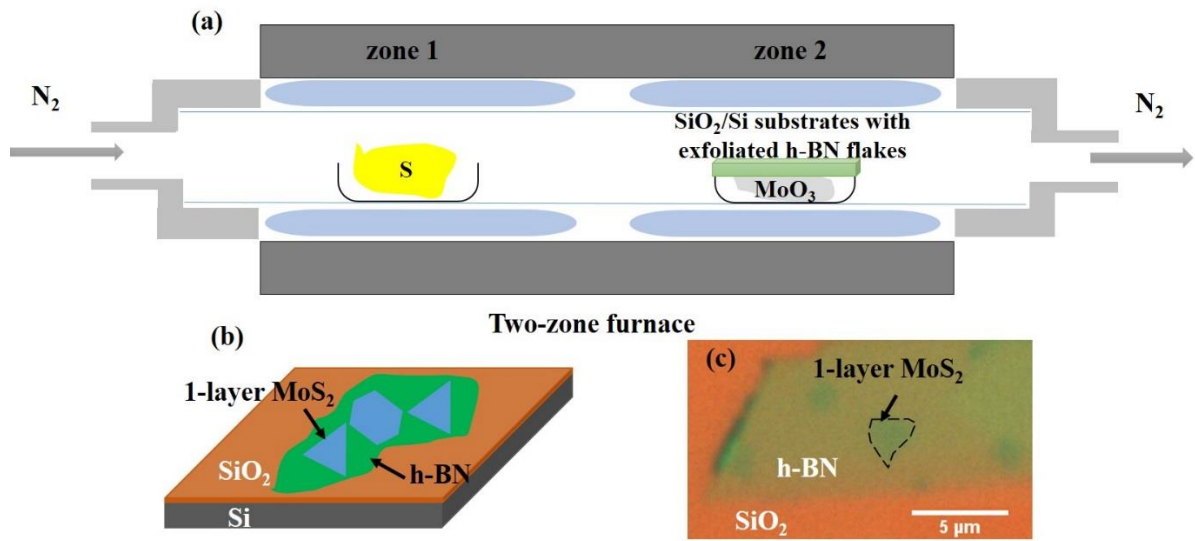



***Figure 1: Experimental setup for the growth, schematic and optical image for representative samples from the growth.*** *(a) Experimental setup for CVD growth of single- and few-layer MoS₂ on exfoliated h-BN. The quartz tube with S and MoO₃ precursors sits in the two-zone furnace. S is in zone 1 while MoO₃ precursor is in zone 2. Exfoliated h-BN flakes on SiO₂/Si chips are placed on top of the crucible that has MoO₃ precursor. N₂ gas runs through the quartz tube during the whole growth process; (b) Schematic of the geometry of as-grown single- and few-layer MoS₂ on exfoliated h-BN flakes, which are on a SiO₂/Si substrate. The green flakes represent thin h-BN (usually less than 200 nm thick) and blue flakes represent MoS₂ that have been grown on h-BN. (c) Optical image of a typical growth of MoS₂ islands on exfoliated thin h-BN flakes on SiO₂/Si substrates. The lighter green flakes are h-BN and the darker green regions are typical MoS₂ flakes. An MoS₂ flake is outlined in (c).*

We examine the atomic-level topography of the $MoS_2$ islands on h-BN via atomic force microscopy (AFM), as shown in Fig. 2. Three distinct $MoS_2$ island topographies are observed: (1) Flat and smooth $MoS_2$ islands (Figs. 2a (ii) and (iii)); (2) Flat and smooth $MoS_2$ islands that surround a tall protrusion at the center (Figs. 2b (ii) and (iii)); and (3) $MoS_2$ islands that exhibit striking helical fringes (Figs. 2c (ii) and (iii)).

We first focus on the (1) and (2) topographies because they are closely related; both represent single-layer $MoS_2$ growth and differ only in the size of the nucleation site. Both (1) and (2) topographies are usually isolated and located randomly on the h-BN flakes. A typical type (1) $MoS_2$ island is outlined with a blue box in Fig. 2a (ii); Fig. 2a (iii) shows a zoom-in of the boxed region. This $MoS_2$ island is ~ 4 $\mu m^2$ in area, which is common for type (1) growth. Often, the $MoS_2$ islands of type (1) are polygon-shaped, while sometimes the grown single-layer $MoS_2$ flake can be strip-like and can be as large as 10 $\mu m^2$ in area (Fig. S1). A line profile from



the edge of the $MoS_2$ island in Fig. 2a (iii) is shown in Fig. 2a (iv). The step edge profile reveals

a height of ~0.7nm, consistent with the height of a single layer of $MoS_2$[2,19]. Although not

revealed in Fig. 2(a), the likely nucleation site for the $MoS_2$ island in type (1) growth is a small

defect in the h-BN, for example a point vacancy[20]. The size and surface inhomogeneity (e.g. step

edges) of h-BN flakes limit the size of single-layer $MoS_2$ that can be grown on h-BN. These two

factors cause the discontinuity of single-layer $MoS_2$ growth in the lateral direction (see examples

shown in Figs. S1 and S2) and potentially cause inhomogeneous nucleation sites for $MoS_2$ to

grow on h-BN.

In Fig. 2b (ii), a flat and smooth $MoS_2$ island of topography (2), which surrounds a tall

protrusion at the center, is outlined with a blue box. Fig. 2b (iii) shows a zoom-in of the boxed

region. $MoS_2$ islands of type (2) are typically flat and smooth with area 1 $\mu m^2$ - 4 $\mu m^2$, and often

have a flower-petal like shape. The tall protrusion in the center is usually smaller than 500 nm

and has a polygon-like shape. Fig. 2b (iv) and its inset shows an AFM line scan, consistent with

single-layer $MoS_2$. The tall pillar-like protrusion in the center of the $MoS_2$ monolayer island is

~25 nm tall and is itself composed of multi-layer $MoS_2$. The pillar likely marks a rather drastic

nucleation site in the underlying h-BN, such as a triangular multi-atom defect[20,21] or impurities

on the surface of h-BN. The growth of the single-layer islands of type (1) and (2) is depicted

schematically in Figs. 2a (i) and 2b (i).

With increased growth time, additional layers of $MoS_2$ can grow on top of the first layer

of type (1) and (2) as discussed above, and form multi-layer $MoS_2$. One example is shown in Fig.

S2, where a smooth tri-layer $MoS_2$ flake without an observable nucleation site is grown on h-BN.

Multi-layer $MoS_2$ obtained in this way follows so-called "layer-by-layer" ("LBL") growth

mechanism.[22] This mechanism is typical for multi-layer $MoS_2$[15,16] and other TMDs[8] grown by



CVD method and is different from the screw-dislocation-driven (SDD) growth mechanism discussed later. An observable nucleation site which causes the type (2) single-layer $MoS_2$ growth sometimes will nucleate multi-layer $MoS_2$ (typically 10 layers or less) over an extended region, much like the extended lower branches of a Christmas tree. An example is provided in Fig. S3, where the as-grown multi-layer $MoS_2$ is at an early stage of forming a Christmas tree shape. The growth of this type of multi-layer $MoS_2$ with observable nucleation sites also follows "LBL" growth mechanism. However, based on our observation, smooth multi-layer $MoS_2$ flakes grown on h-BN rarely show observable nucleation sites.

Fig. 2c (ii) shows an example of topography (3), which is distinct from topographies (1) and (2): the $MoS_2$ islands are pyramid-like with hexagonal or triangular bases. The type (3) islands are of different maximum thickness and are located randomly on the h-BN flakes. Interestingly, the type (3) islands have a helical (spiral) structure in the normal direction. Fig. 2c (iii) shows a zoom-in image of one of the islands, and Fig. 2c (iv) shows the result of an AFM line scan acquired along the red line of Fig. 2c (iii)). The entire island structure identified here has a height of ~10 nm and step-like features with 0.86 nm heights. This clearly represents multi-layer $MoS_2$, grown in a screw-like manner.

Topography (3) multilayer $MoS_2$ islands result from a screw-dislocation-driven (SDD) growth mechanism. This has also been observed in other CVD grown TMD materials [23,24] and has been attributed to a low supersaturation condition. The SDD growth mechanism is also a common growth mode observed in other anisotropic nanostructure growths[25–28]. We depict this growth in Figure 2c (i) as a cross-sectional schematic of multilayer $MoS_2$ grown on h-BN. For SDD growth of a 2-D material, the starting point is typically a vertical offset (or slip) in the atomic planes of the first growth layer. Once the screw dislocation is created, the following



layers tend to nucleate and grow from the exposed edge of the dislocation due to the decreased energy barrier at these sites, which promotes more vertical growth than lateral growth. Both the helical features and the profile with a step height of 0.86 nm, which is close to the thickness of single-layer $MoS_2$, indicates the type (3) multi-layer islands of $MoS_2$ represent SDD growth with a single elementary Burgers vector for the screw dislocation[23,29]. We have also observed herringbone contours (Fig. S4) in few-layer $MoS_2$ grown on h-BN, which are typical features in SDD growth[23,29]. An identifying feature of this growth mechanism is an extended offset in height corresponding to the Burger's vector of the screw dislocation (or a slipped edge).[23,24,28] Because our AFM scans of as-grown single-layer $MoS_2$ on h-BN did not show this identifying feature, we conclude the SDD growth mechanism does not produce single-layer $MoS_2$ in our case.

Here we summarize the growth mechanisms for single-layer and multi-layer $MoS_2$ grown on h-BN. Smooth single-layer $MoS_2$ can grow on h-BN with or without observable nucleation sites, while smooth multi-layer $MoS_2$ can grow on h-BN following "LBL" growth mechanism, which is an extension of smooth single-layer $MoS_2$ growth. Multi-layer $MoS_2$ can also nucleate and grow on h-BN from a screw dislocation, which is called SDD growth mechanism.



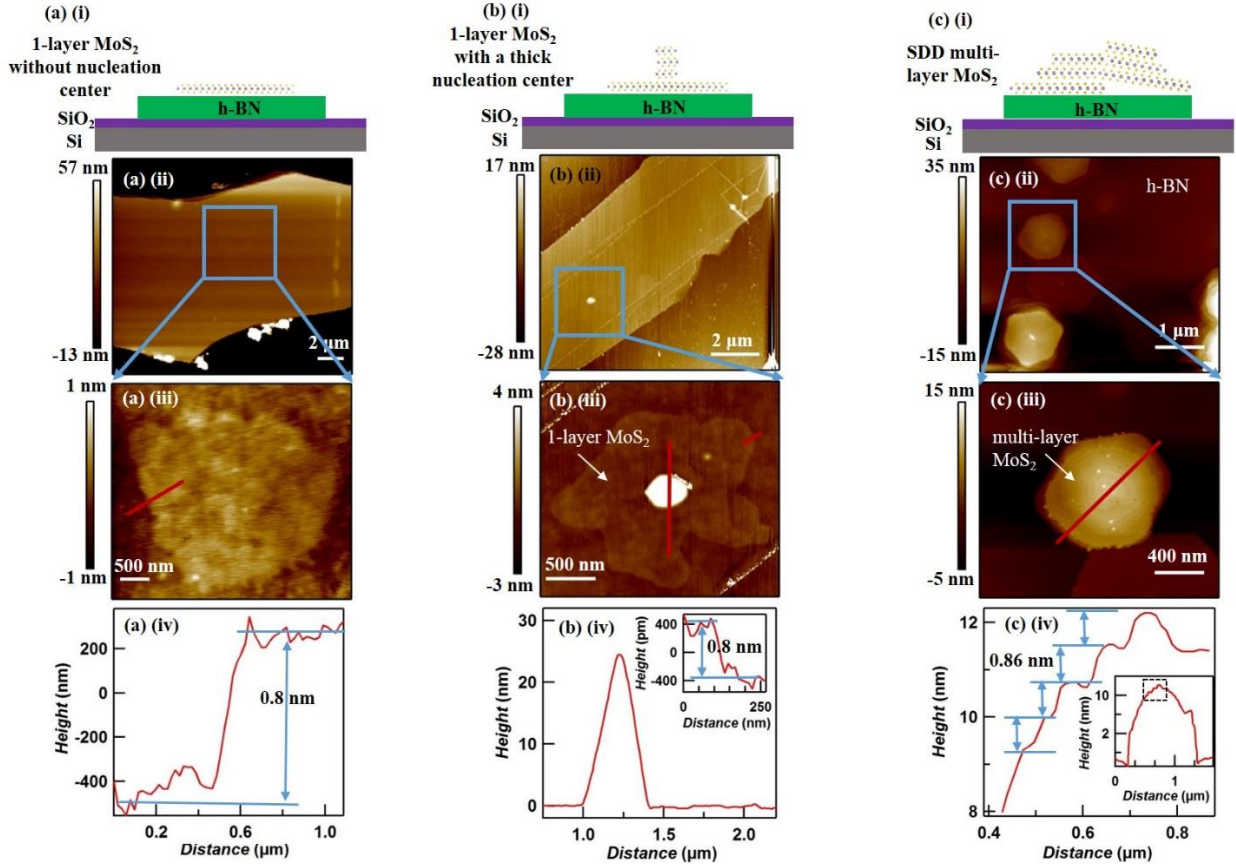

***Figure 2: AFM characterization of single-layer and multi-layer MoS₂ grown on exfoliated h-BN, which shows different growth mechanisms of MoS₂ on exfoliated h-BN by CVD method.*** *(a)(b) both show the typical geometry of single-layer MoS₂ grown on h-BN. (a) shows a single-layer MoS₂ without a nucleation center can grow on h-BN. (a) (i) is the schematic of such growth. The green slab represents an h-BN flake and the sandwich-structured MoS₂ is grown on top. (a) (ii) is the low-magnification AFM image of single-layer MoS₂ on h-BN and (a) (iii) is the zoom-in image of the area outlined by the blue box in a(ii). The height profile at the edge of MoS₂ shows the thickness is around 0.7 nm, which is consistent with single-layer MoS₂. (b) (i) shows the schematic for single-layer MoS₂ with a thick nucleation center. (b) (ii) shows the low-magnification AFM image of a single-layer MoS₂ with a thick nucleation center on h-BN and b(iii) is the zoom-in image of the MoS₂ flake outlined in b(ii). b(iii) is the height profile across*



the nucleation center and the edge of $MoS_2$, showing the nucleation center is around 25 nm, while the edge shows a thickness of a single-layer $MoS_2$. c(i) is the schematic for multi-layer $MoS_2$ grown on h-BN. A typical multi-layer $MoS_2$ island grown on h-BN follows the screw-dislocation-driven (SDD) growth mechanism. c(i) shows the growth starts from a screw-dislocation created at the interface of two $MoS_2$ flakes with one elementary burgers vector displaced vertically. c(ii) is the low-magnification AFM image of a few multi-layer $MoS_2$ islands grown on h-BN. c(iii) shows the zoom-in image of one $MoS_2$ island outlined in c(ii). The height profile across the center of $MoS_2$ island in c(iv) shows the step size of ~0.86 nm, which is about the thickness of one-layer $MoS_2$. The color scale for all the AFM images is adjusted so that as-grown $MoS_2$ flakes can be visualized from the contrast.

We employ transmission electron microscopy (TEM) to further characterize $MoS_2$ grown on h-BN. Fig. 3a shows for a type (2) multi-layer $MoS_2$ island the high-angle-annular-dark-field (HAADF) image. Since $MoS_2$ is atomically heavier than h-BN the HAADF image will show significant contrast between $MoS_2$ and h-BN. Indeed, in Fig. 3a we observe a bright triangular area on top of a distinct dark background, representing the presence of $MoS_2$ on h-BN. Energy-dispersive X-ray spectroscopy (EDS) mapping with distributions of Mo, S, B and N are also shown in Fig. 3(b-e), respectively. In this EDS mapping, B and N are found in the entire area indicating h-BN is present everywhere within the observation window, as expected. Mo and S are distributed in a manner similar to the shape of the bright contrast in Fig. 3a, and are clearly attributed to $MoS_2$. EDS analysis also allows the Mo:S ratio to be determined; we find 34.9±3.5 to 65.1±2.0, consistent (considering experimental uncertainties) with the expected composition 1:2 for $MoS_2$. Our TEM EDS measurements thus unambiguously confirm the flakes grown in this study as $MoS_2$.



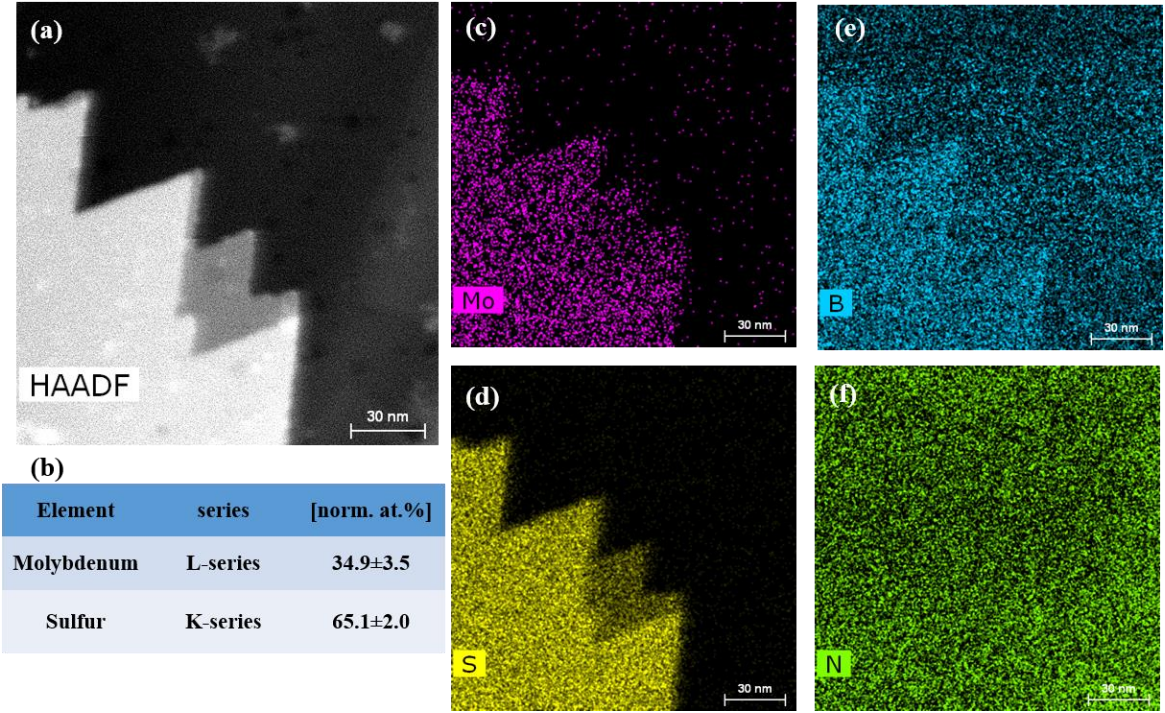

The image panels contain the following labels: (a) HAADF, 30 nm; (c) Mo, 30 nm; (e) B, 30 nm; (d) S, 30 nm; (f) N, 30 nm.

(b)

| Element | series | [norm. at.%] |
|---|---|---|
| Molybdenum | L-series | 34.9±3.5 |
| Sulfur | K-series | 65.1±2.0 |

***Figure 3: STEM EDS mapping of thin MoS$_2$ grown on exfoliated h-BN, showing the composition/stoichiometry of MoS$_2$ flakes on h-BN.*** *(a) High-angle-annular-dark-field (HAADF) image of few-layer MoS$_2$ on h-BN. The white region is as-grown thin MoS$_2$ while the dark background is h-BN. (c)(d)(e)(f) are the elemental maps of MoS$_2$ on h-BN, showing the location of Mo, S, B and N respectively. The Mo and S maps clearly show the MoS$_2$ flake. N map shows uniform distribution while B map shows higher intensity around the location of the MoS$_2$ flake. The non-uniform distribution of B is due to the overlap of the B and Mo peaks in the spectrum (Fig. S5): the concentration of B is based on the intensity of the B peak around 0.18 keV and this B peak may include some intensity from the Mo peak located around the same energy. (b) provides the atomic composition of Mo (34.9±3.5%) and S (65.1±2.0%), analyzed from the STEM EDS mapping. The atomic ratio between Mo and S is close to the expected 1:2 for MoS$_2$.*



To characterize the quality of monolayer $MoS_2$ crystals grown on h-BN, we perform photoluminescence (PL) experiments (Fig. 4), and compare the results to PL measurements on single-layer $MoS_2$ grown on $SiO_2$ via CVD. AFM is used to ensure that the $MoS_2$ samples are single layer. Our single-layer $MoS_2$ grown on h-BN has a strong PL peak centered at 1.89 eV (Fig. 4). This measured direct band gap is quite close to the one of free-standing exfoliated single-layer $MoS_2$- 1.90 eV[30], and is larger than the CVD grown $MoS_2$ on $SiO_2$ (1.84eV[15]) and exfoliated single-layer $MoS_2$ on $SiO_2$ (1.85eV[31]). The full width at half maximum (FWHM) of the PL peak from as-grown $MoS_2$/h-BN heterostructure is approximately 40 meV, which is slightly smaller than that for CVD grown $MoS_2$ on $SiO_2$ (50 meV) as shown in Fig. 4 and also in reference[15]) and free-standing exfoliated $MoS_2$ (50-60 meV[30]), and is much smaller than $MoS_2$ exfoliated onto $SiO_2$ (100- 150 meV[31]). These characteristics of PL indicate that $MoS_2$ grown on h-BN is electronically less perturbed than that grown on $SiO_2$, and is more like free-standing $MoS_2$. We note, the high growth temperature can cause stretching of as-grown $MoS_2$ after the sample cools down to room temperature, due to h-BN's negative lateral thermal expansion coefficient and $MoS_2$'s positive lateral thermal expansion coefficient. The photoluminescence peak from stretched single-layer $MoS_2$ will red-shift compared to single-layer $MoS_2$ grown on $SiO_2$.[32] However, the photoluminescence peak from single-layer $MoS_2$ grown on h-BN studied here was consistently observed to be close to the peak from free-standing monolayer $MoS_2$, and blue-shifted compared to $MoS_2$ grown on $SiO_2$. Thus, the effect of the electrical environment for $MoS_2$ grown on h-BN is greater than the influence from strain effects. This supports our claim that $MoS_2$ grown on h-BN is less electrically disturbed, which is consistent with studies on single-layer $MoS_2$ transferred onto h-BN.[13,14]



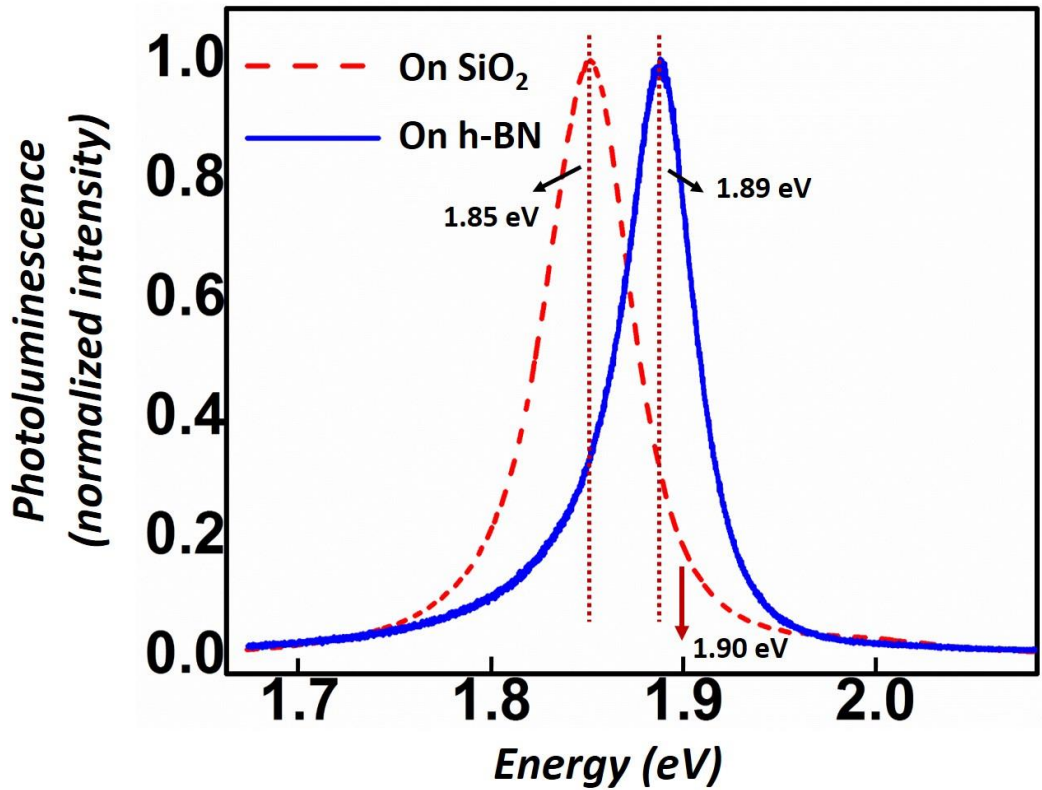

**Figure 4: Photoluminescence from single-layer MoS₂ grown on h-BN.** *Photoluminescence peak from single-layer MoS₂ grown on h-BN (1.89 eV) indicates a band gap closer to free-standing MoS₂ flake (1.90 eV, pointed out by the arrow) compared to that grown on SiO₂ (1.85 eV).*

We now turn to the relative crystal orientation of MoS₂ grown on h-BN. The relative rotation angle between the constituent layers in a 2-D heterostructure can play a significant role in the electronic band structure of the heterostructure.[7,33,34] Although MoS₂ and h-BN both exhibit hexagonal crystal structure, the associated lattice constants differ by >20% and are incommensurate; hence it is *a priori* unclear if any preferred orientation between MoS₂ and h-BN lattices should occur. We find, however, that there is a strong orientation preference.



We examine a dilute collection of independent, nearly identical small (~500nm) triangular crystallites of $MoS_2$ grown on single-crystal h-BN, and observe two highly dominant relative orientations, differing by 60° (Fig. S7). We use selected area electron diffraction (SAED) in TEM to confirm the absolute orientation angles: the peaks in the bimodal distribution indeed correspond to crystal lattices' alignment, with the orientation of "S" sublattice of $MoS_2$ aligned with either the "B" sublattice or the "N" sublattice of h-BN, with equal probability. In other words, the orientation of $MoS_2$ is sensitive to the atomic corrugation of h-BN, but it does not appreciably distinguish between B and N atoms. In our detailed SAED analysis below, we exploit this indistinguishability and conveniently employ a reduced angle definition for the relative orientation angle (see Fig. S6), where the reduced misorientation angle spans 0 to 30 degrees.

We focus for the moment on single-layer $MoS_2$ grown on h-BN and employ SAED to characterize the rotation angle for 22 $MoS_2$ overlayers. Fig. 5a shows a typical SAED pattern in which there are two sets of six-fold symmetric diffraction spots. The six spots of the inner hexagon (denoted by green lines) correspond to $MoS_2$ ($a_{MoS_2}= 3.1$Å) and the six spots of the outer hexagon (denoted by purple lines) correspond to h-BN ($a_{h-BN}=2.5$ Å). From this pattern we measure a relative rotation angle between $MoS_2$ and h-BN of ~9°. By selecting one diffraction spot of $MoS_2$, one can visualize the $MoS_2$ flake in dark field image (Figure 5b), where $MoS_2$ appears bright and the dark background is h-BN. Based on the area probability histogram for specific relative rotation angles between the as-grown single-layer $MoS_2$ and h-BN (from 22 locations), a low angle (< 6°) is most dominant (around 45% area fraction), as shown in Fig. 5(c). The histogram for counts also shows that 10 out of 22 single-layer $MoS_2$ flakes grown on h-



BN have relative rotation angles < 6° (Figure 5d). Single-layer $MoS_2$ flakes that have a relative rotation angle 6°-12° and 24°- 30° are also present but are less prevalent.

The preferred low relative rotation angles between $MoS_2$ and h-BN can be attributed to van der Waals epitaxy[36] that is modified by several factors. Van der Waals epitaxy permits one type of 2-D material to grow on another type in a rotationally commensurate manner, despite the highly mismatched lattice constants of the constituent materials. The slightly broadened distribution in rotation angle (within 6° range) in our $MoS_2$/h-BN heterostructure suggests additional factors play a role in deviation from van der Waals epitaxy. This has also been observed in previous studies of other directly grown TMDs on various substrates.[37,38]

The study of relative rotation angle between multi-layer $MoS_2$ and the h-BN substrate is less straightforward due to the more complex growth mechanism of multi-layer $MoS_2$ islands grown on h-BN compared to the single-layer $MoS_2$ case. In this case we often observe multiple relative rotation angles although the h-BN substrate may be one single-crystal domain. Fig. 5(e) shows a triangular $MoS_2$ thick island grown on h-BN. The significant contrast at the center of the $MoS_2$ flake is caused by the screw-dislocation. The SAED pattern taken from the outlined area in Fig. 5(e) is shown in Fig. 5(f). There are at least two relative rotation angles in Fig. 5(f), but one strongly apparent and symmetric set of diffraction spots from $MoS_2$ shows the relative rotation angle is 3°, which is consistent with the most probable relative rotation angle for single-layer



MoS$_2$ grown on h-BN as shown in Fig. 5(c and d).

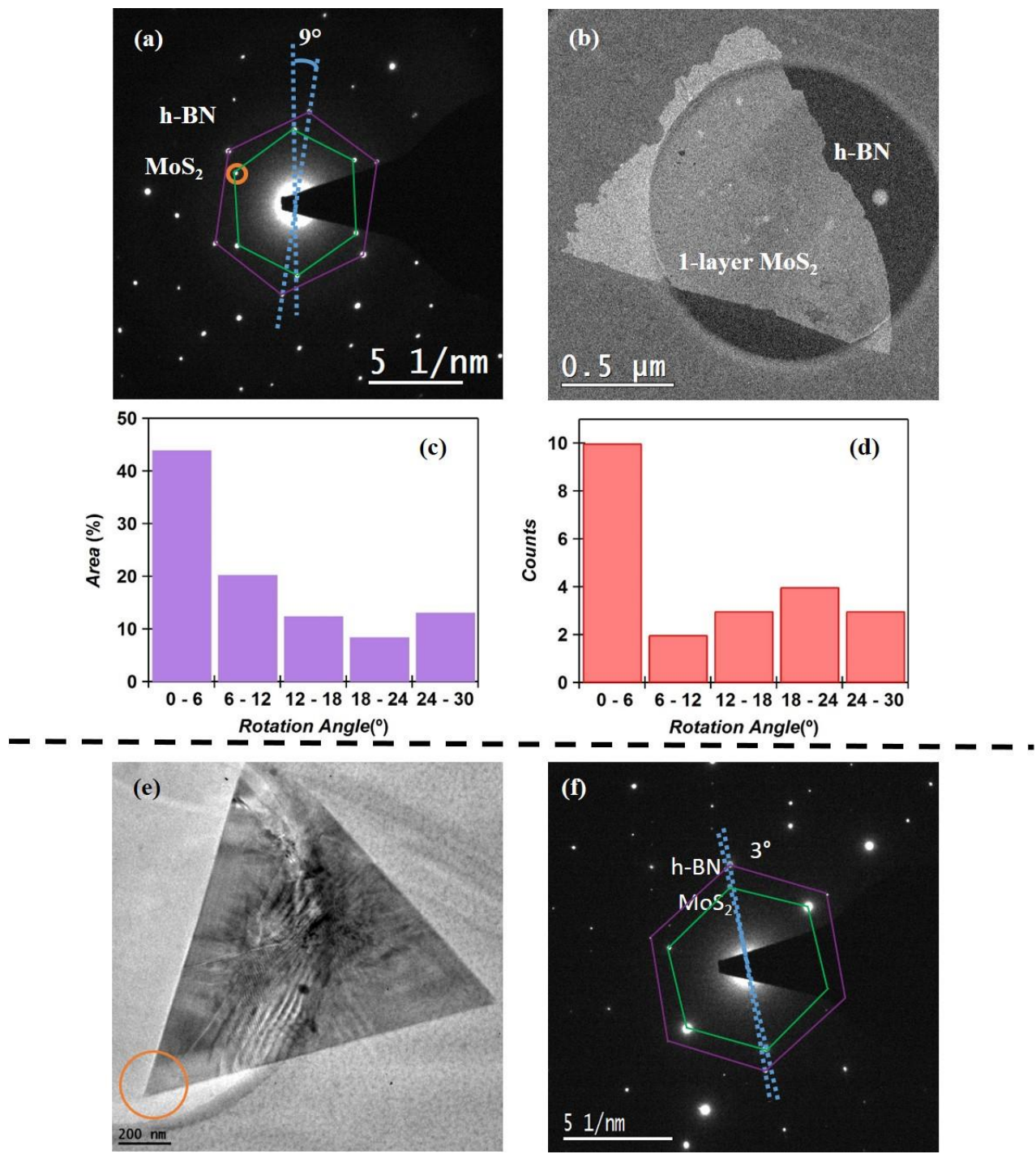

***Figure 5: TEM characterization of single-layer and multi-layer MoS$_2$ grown on exfoliated h-BN.*** *(a) A typical selected area electron diffraction (SAED) pattern of a single-layer MoS$_2$ flake*



*grown on thin h-BN. The green hexagon shows the six-fold-symmetric diffraction spots from MoS$_2$ while the purple hexagon shows the six-fold-symmetric diffraction spots from h-BN. The relative rotation angle between MoS$_2$ and h-BN is measured to be 9° from this diffraction pattern. By selecting one of the diffraction spots from MoS$_2$ (outlined in (a)), one can visualize the MoS$_2$ flake in dark filed image shown in (b). The white region in (b) is the single-layer MoS$_2$ while h-BN appears dark. The white dot visible on h-BN is residual polymer from TEM sample preparation, which also indicates the dark region is not empty. The heterostructure of MoS$_2$ and h-BN lies over a hole on a quantifoil TEM grid. (c) shows the area probability histogram of the relative rotation angle of single-layer MoS$_2$ grown on h-BN based on 22 locations of such growth. (d) shows the count histogram of the same growths in (c). (e) is a TEM image of a typical triangular multi-layer MoS$_2$ island grown on h-BN. The contrast due to the screw dislocation is visible around the center of MoS$_2$ island. (f) is the diffraction pattern from the region outlined in (d). The green hexagon outlines the diffraction spots from MoS$_2$ while the purple hexagon outlines the diffraction spots from h-BN. The relative rotation angle between MoS$_2$ and h-BN at this specific location is ~3°.*

In summary, we have demonstrated that single-layer and few-layer MoS$_2$ can be directly grown on h-BN by CVD method. The growth mechanisms were found to differ depending on the different supersaturation condition of precursors. Under low supersaturation condition, screw-dislocation-driven growth dominates and causes the few-layer MoS$_2$ to form striking helical structures. Otherwise, smooth single-layer and few-layer MoS$_2$ with non-observable or observable nucleation centers can form on h-BN, following "layer-by-layer" growth mechanism. The as-grown single-layer MoS$_2$ on h-BN shows a strong photoluminescence peak centered around 1.89 eV, which is closer to that of free-standing MoS$_2$. This indicates single-layer MoS$_2$



grown on h-BN has less perturbed electrical environment and is promising for high-quality

MoS$_2$-based devices. Detailed TEM studies show that single-layer MoS$_2$ grown on h-BN has

preferred low relative rotation angle between the two, which is also interesting for further study

of electronic band structure modification due to different relative rotation angle in this

heterostructure.

ASSOCIATED CONTENT

**Supporting Information:** Including "Materials and methods", Figures S1 to S7. This material is

available free of charge via the internet at http://pubs.acs.org.

AUTHOR INFORMATION


**Corresponding Author**

[*]To whom correspondence should be addressed:  azettl@berkeley.edu



ACKNOWLEDGEMENT:

This research was supported in part by the Director, Office of Basic Energy Sciences, Materials

Sciences and Engineering Division, of the U.S. Department of Energy under Contract DE-AC02-

05CH11231, within the sp2-bonded Materials Program, which provided for postdoctoral support

and AFM and PL characterization by NSF grant DMR-1206512 which provided for the sample

growth; and by the Molecular Foundry of the Lawrence Berkeley National Laboratory, under

Contract DE-AC02-05CH11231, which provided for TEM characterization. We acknowledge

Karen Bustillo in the Molecular Foundry of the Lawrence Berkeley National Laboratory for

TEM technical support and Dr. Wei Chen in the Electrochemical Technologies Group of the




Lawrence Berkeley National Laboratory for useful discussion. Illuminating discussions with Ashley Gibb are also acknowledged.

*Note added*- After submission of this work, we became aware of a related independent report[39] of $MoS_2$ grown directly on CVD prepared h-BN (which had been transferred to $SiO_2$ substrates for $MoS_2$ growth).

**TOC figure:**

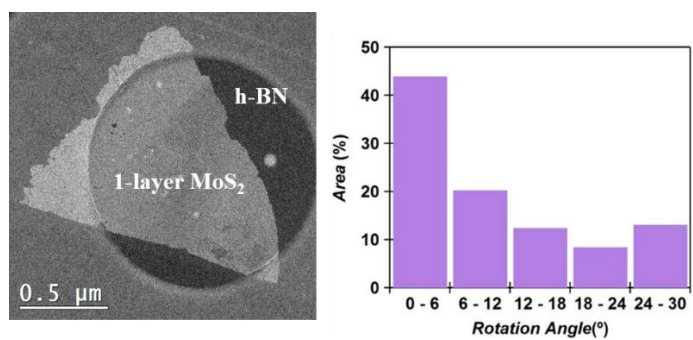